\documentclass[aps,nofootinbib,twocolumn, showpacs]{revtex4-1}

\usepackage[utf8]{inputenc}
\usepackage{amssymb}
\usepackage{amsmath}
\usepackage{amsfonts}
\usepackage{dsfont}
\usepackage[usenames,dvipsnames]{xcolor}
\usepackage[pdftex]{graphicx}
\usepackage[pdftex]{hyperref}
\hypersetup{plainpages=false,colorlinks=true,linkcolor=Red, citecolor=blue, urlcolor=blue}
\usepackage{bm}

\usepackage{array}
\newcolumntype{L}[1]{>{\raggedright\let\newline\\\arraybackslash\hspace{0pt}}m{#1}}
\newcolumntype{C}[1]{>{\centering\let\newline\\\arraybackslash\hspace{0pt}}m{#1}}
\newcolumntype{R}[1]{>{\raggedleft\let\newline\\\arraybackslash\hspace{0pt}}m{#1}}

\renewcommand{\vec}[1]{\bm{\mathrm{#1}}}

\begin{document}

\title{Phenomenological theories of the low-temperature pseudogap:\\Hall number, specific heat and Seebeck coefficient}
\author{S. \surname{Verret}}
\email[Corresponding author: ]{simon.verret@usherbrooke.ca}
\affiliation{D\'epartement de physique, Universit\'e de Sherbrooke, Qu\'ebec, Canada  J1K 2R1}
\author{O. \surname{Simard}}
\affiliation{D\'epartement de physique, Universit\'e de Sherbrooke, Qu\'ebec, Canada  J1K 2R1}
\author{M. \surname{Charlebois}}
\affiliation{D\'epartement de physique, Universit\'e de Sherbrooke, Qu\'ebec, Canada  J1K 2R1}
\author{D. \surname{S\'en\'echal}}
\affiliation{D\'epartement de physique, Universit\'e de Sherbrooke, Qu\'ebec, Canada  J1K 2R1}
\author{A.-M. S. \surname{Tremblay}}
\affiliation{D\'epartement de physique, Universit\'e de Sherbrooke, Qu\'ebec, Canada  J1K 2R1}
\affiliation{Canadian Institute for Advanced Research, Toronto, Ontario, Canada M5G 1Z8}
\date{\today}
\begin{abstract}
Since its experimental discovery, many phenomenological theories successfully reproduced the rapid rise of the Hall number $n_H$, going from $p$ at low doping to $1+p$ at the critical doping $p^*$ of the pseudogap in superconducting cuprates. Further comparison with experiments is now needed in order to narrow down candidates.
In this paper, we consider three previously successful phenomenological theories in a unified formalism---an antiferromagnetic mean field (AF), a spiral incommensurate antiferromagnetic mean field (sAF), and the Yang-Rice-Zhang (YRZ) theory. We find a rapid rise in the specific heat and a rapid drop in the Seebeck coefficient for increasing doping across the transition in each of those models. The predicted rises and drops are locked, not to~$p^*$, but to the doping where anti-nodal electron pockets, characteristic of each model, appear at the Fermi surface shortly before~$p^*$. While such electron pockets are still to be found in experiments, we discuss how they could provide distinctive signatures for each model. We also show that the range of doping where those electron pockets would be found is strongly affected by the position of the van~Hove singularity. 

\end{abstract}

\maketitle

\section{Introduction}

In clean YBCO crystals, the Hall conductivity $R_H$ undergoes an abrupt change at doping~\mbox{$p^*\sim 0.19$}. In the low-temperature magnetic-field induced normal state, the Hall number $n_H=1/eR_H$ rises rapidly from $n_H=p$ at at the end of the charge-ordered phase, to $n_H=1+p$, expected for the Fermi liquid regime at high doping \cite{badoux_change_2016}; a loss of carriers below $p^*$ is conjectured to be the cause. This discovery received much attention, and many theoretical models were shown to reproduce this behavior: an antiferromagnetic (AF) mean field~ \cite{storey_hall_2016}, a spiral antiferromagnetic (sAF) mean field~ \cite{eberlein_fermi_2016}, the Yang-Rice-Zhang (YRZ) theory~ \cite{yang_phenomenological_2006,storey_hall_2016}, a $\mathds{Z}_2$ fractionalized Fermi-liquid~(FL$^*$)~theory \cite{chatterjee_fractionalized_2016}, a nematic transition \cite{maharaj_hall_2016}, a $SU(2)$ fluctuation model \cite{morice_evolution_2017}. All of the above successfully reproduce the rapid rise in Hall number because they entail changes in the Fermi surface at $p^*$. They are all phenomenological theories in which those changes were set up to happen precisely at~$p^*$. 
Additional work on a unidirectional charge density wave model~\cite{sharma_suppression_2017}, and an incommensurate collinear spin density wave model~\cite{charlebois_hall_nodate} also provide insight on the matter.
To isolate the strengths and weaknesses of all the above models, more comparison with experiments is needed. This paper makes verifiable predictions for three of the above models.

The Hall effect is not the only probe capable of studying the changes happening at $p^*$. 
Recent resistivity measurements were argued to account for the same loss in carrier density \cite{collignon_fermi-surface_2017}, with theoretical investigations arriving at similar conclusions \cite{chatterjee_thermal_2017,moutenet_phenomenological_nodate}.
Regarding earlier studies, specific heat $C_v$ measurements provided evidence that the low temperature density of states increases significantly from underdoped samples to overdoped samples \cite{loram_condensation_2000,loram_specific_1998,momono_low-temperature_1994}, consistent with a gap closing as doping increases~\cite{yoshida_low-energy_2007}. 
As a consequence, a corresponding increase should be observable in $C_v/T$ at low temperature, as a function of doping.
Similarly, the finite temperature Seebeck $S_x$ coefficient decreases significantly from underdoped to overdoped samples \cite{mcintosh_van_1996,kondo_contribution_2005, munakata_thermoelectric_1992, fujii_-plane_2002,daou_thermopower_2009}.
As $S_x$ must vanish at zero temperature, the corresponding increase should be observable in $S_x/T$ at low temperature as a function of doping.
Therefore, under the same experimental conditions as for the Hall number \cite{badoux_change_2016}---low temperature with superconductivity suppressed by a high magnetic field---we expect that measurements of the specific heat and Seebeck coefficient will provide clarifications on the nature of the $p^*$ transition. 
To our knowledge, however, no such normal state data for the specific heat nor the Seebeck effect at low temperature as function of doping is available in the literature.

In this paper, we compute the predictions for the low temperature normal state electronic specific heat~$C_{V}/T$ and the Seebeck coefficient~$S_x/T$ as a function of doping, for the antiferromagnetic (AF), incommensurate spiral antiferromagnetic (sAF), and Yang-Rice-Zhang (YRZ) models mentioned above.
Thus, we extend the results of Refs.~\onlinecite{storey_hall_2016,eberlein_fermi_2016} for the Hall number and those of Refs.~\onlinecite{leblanc_specific_2009, storey_electron_2013} for the specific heat and Seebeck effect in YRZ theory.
All three models are compared in a unified formalism.
We also study the effects of band structure, notably the role of the van~Hove singularity and its proximity to $p^*$. Finally, we compare how the two limiting cases of isotropic mean-free path~$\ell$ and constant lifetime~$\tau$ approximations affect all computations. 

Our main result consists of a rapid rise in $C_V/T$ and a drop in $S_x/T$ when increasing doping across $p^*$, at low temperature.
In all three models, characteristic electron pockets appear at the Fermi surface when approaching~$p^*$.\footnote{Although electron pockets appearing at the Fermi surface constitute what is commonly known as a ``van Hove singularity'', in this paper we use ``van Hove singularity'' for when the Fermi level crosses the saddle point of the dispersion. The former is simply called ``when electron pockets appear at the Fermi surface''.}
The rise and drop found in $C_V/T$ and $S_x/T$, respectively, are not located at~$p^*$, but rather at the lower doping $p_e$ where these electron pockets appear.
This result extends the conclusion by Storey~\cite{storey_hall_2016}, that the width of the rise in Hall number is entirely determined by the range of doping occupied by these anti-nodal electron pockets. For the Seebeck effect and specific heat, almost no signature appears at~$p^*$, as if the electron pockets displaced the transition. 

The paper is divided as follows. Section~\ref{section:models} presents the three separate starting points of the AF, sAF and YRZ models. Section~\ref{section:methods} presents the unified formalism used to treat all three models.
Section~\ref{section:results} discusses results. Finally, section~\ref{section:conclusion} highlights our main conclusions, and appendix~\ref{appendix:scattering} provides a brief analysis of bare band results across the van Hove singularity (vHs) for the constant lifetime~$\tau$ and isotropic mean-free-path~$\ell$ approximations.

\section{Models}
\label{section:models}
\noindent
We start with the one-band tight-binding dispersion:
\begin{align}
\xi_{\vec k} &=
-2t (\cos(k_xa)+\cos(k_ya))
\nonumber\\&\quad- 2t'(\cos(k_xa+k_ya)+\cos(k_xa-k_ya))
\nonumber\\&\quad- 2t''(\cos(2k_xa)+\cos(2k_ya))-\mu.
\label{band}
\end{align}
All energies are measured relative to the first neighbor hopping amplitude (one can set $t=250$~meV for comparison with experiments), $\mu$ is the chemical potential, $\hbar\vec k$ is the crystal momentum, $a$ the lattice spacing and we study various sets of band parameters $t'$ and $t''$, corresponding to the second and third neighbor hopping, respectively. The values used are indicated in the corresponding figures.

\subsection{Antiferromagnetism}
\noindent
The AF model is defined by the Hamiltonian:
\begin{align}
H^{\text{AF}}&=\sum_{\vec k}
\begin{pmatrix}c^\dagger_{\vec k\uparrow}&c^\dagger_{\vec k+\vec Q,\uparrow}\end{pmatrix}
\begin{pmatrix}
\xi_{\vec k} & \Delta \\
\Delta &\xi_{\vec k+\vec Q}
\end{pmatrix}
\begin{pmatrix}c^{\phantom{\dagger}}_{\vec k,\uparrow}
\\c^{\phantom{\dagger}}_{\vec k+\vec Q,\uparrow}\end{pmatrix},
\label{af_hamiltonian}
\end{align}
where $\vec Q = (\pi, \pi)$ is the antiferromagnetic wave vector, $c^\dagger_{\vec k\uparrow}$ is the operator that creates a Bloch electron of momentum~$\hbar\vec k$ and spin up. 
We only write the Hamiltonian for spin up because the only difference for spin down is the sign of the gap energy $-\Delta$. Since both gap signs lead to the same eigenvalues, spin up and down are equivalent in this model. As a consequence, instead of working in the reduced AF Brillouin zone and multiplying everything by 2 for spin, we ignore this explicit factor 2 and work in the original Brillouin zone. This helps to unify the three models in section~\ref{section:methods}.


\subsection{Incommensurate spiral antiferromagnetism}
\noindent
The sAF model is defined by the Hamiltonian:
\begin{align}
H^{\text{sAF}}&=\sum_{\vec k} 
\begin{pmatrix}c^\dagger_{\vec k,\uparrow}&c^\dagger_{\vec k+\vec Q,\downarrow}\end{pmatrix}
\begin{pmatrix}
\xi_{\vec k} & \Delta\\
\Delta &\xi_{\vec k+\vec Q}
\end{pmatrix}
\begin{pmatrix}c^{\phantom{\dagger}}_{\vec k,\uparrow}
\\c^{\phantom{\dagger}}_{\vec k+\vec Q,\downarrow}\end{pmatrix}.
\label{saf_hamiltonian}
\end{align}
In that case, $\vec Q$ is incommensurate and changes with doping following $\vec Q(p) = (\pi-2\pi p, \pi)$ \cite{eberlein_fermi_2016,chatterjee_thermal_2017}. The sum on $\vec k$ spans the original Brillouin zone. Here the sAF order parameter $\Delta$ couples spin up with spin down. Consequently, the two spin contributions to transport must be computed separately.

The only fundamental difference between the AF and the sAF models is the $\vec Q$ vector. 
Using $\vec Q=(\pi,\pi)$ in Hamiltonian~\eqref{saf_hamiltonian} would lead to antiferromagnetism perpendicular to the spin quantization axis with the same eigenvalues as Hamiltonian~\eqref{af_hamiltonian} and hence the same transport results.
In the unified formalism of section~\ref{section:methods}, the additional differences appearing simply outline two ways of getting the same thing: the AF model could also be formulated as a sAF model with commensurate $\vec Q=(\pi,\pi)$. However, the converse is not true: the sAF model cannot be expressed as the AF model with incommensurate $\vec Q(p) = (\pi-2\pi p, \pi)$.

\subsection{Yang-Rice-Zhang theory}
\noindent
Contrary to the AF and sAF models, YRZ theory \cite{yang_phenomenological_2006} is defined not from a Hamiltonian, but from the following Green's function ansatz, valid for both spins:
\begin{align}
G^{\text{YRZ}}_{\vec k}(\omega)\equiv
\frac{g_t(p)}{\omega-\xi^{g}_{\vec k}(p)-\displaystyle{\frac{|\Delta^{\text{PG}}_{\vec k}(p)|^2}{\omega-\xi^{0}_{\vec k}(p)}}} + G_{\text{inc.}}.
\label{yrz_ansatz}
\end{align}
This ansatz uses the renormalized dispersion: 
\begin{align}
\xi^{g}_{\vec k}(p) &=
-(g_{t}(p) + \tfrac{3\chi J}{8}g_{s}(p) )\cdot 2t (\cos(k_xa)+\cos(k_ya))
\nonumber\\&\quad- g_{t}(p)\cdot 2t' (\cos(k_xa+k_ya)+\cos(k_xa-k_ya))
\nonumber\\&\quad- g_{t}(p)\cdot2t''(\cos(2k_xa)+\cos(2k_ya))-\mu,
\label{band_gutz}
\end{align}
with $g_t(p)=\tfrac{2p}{1+p}$ and $\frac{3\chi J}{8} g_s(p)=\tfrac{0.169}{(1+p)^2}$ being standard Gutzwiller factors that flatten the band as a function of doping $p$.
The role of these factors is to approximate the loss of metallicity when approaching the Mott insulator \cite{leblanc_signatures_2014}. Therefore, whereas Eq.~\eqref{band} represents a non-interacting band, Eq.~\eqref{band_gutz} represents the renormalized band expected from a doped Mott insulator.

The third term in the denominator of Eq.~\eqref{yrz_ansatz}, the self-energy, relies on another dispersion:
\begin{align}
\xi^{0}_{\vec k}(p)&=2t (g_{t}(p) + \tfrac{3\chi J}{8}g_{s}(p) )(\cos(k_x)+\cos(k_y)).
\label{yrz_disp}
\end{align}
This dispersion corresponds exactly to the first term of~$-\xi^{g}_{\vec k}(p)$ and the resulting pseudogap opens on the so-called umklapp surface at the core of YRZ theory. Note that $\xi^{0}_{\vec k}(p)$ also corresponds to the first term of $\xi^{g}_{\vec k+(\pi,\pi)}(p)$, and the umklapp surface corresponds to the AF zone boundary, explaining the strong resemblance with the AF model. In this respect, $\xi^{0}_{\vec k}(p)$ can be seen as the dispersion of ancillary excitations with perfect $\vec Q=(\pi,\pi)$ susceptibility \cite{leblanc_signatures_2014}.

YRZ theory couples those two dispersions, $\xi^{g}_{\vec k}(p)$ and $\xi^{0}_{\vec k}(p)$, with a $d$-wave pseudogap order parameter decreasing monotonically with doping in the range \mbox{$0<p<0.2$}:
\begin{align}
\Delta^{\text{PG}}_{\vec k}(p)=\frac{3t}{2}(\cos(k_x)-\cos(k_y))(0.2-p).
\end{align}
The order parameter's maximum amplitude is at the antinodes, with \mbox{$\Delta^{\text{PG}}_{(0,\pi)}=-\Delta^{\text{PG}}_{(\pi,0)}=3t(0.2-p)$}. Note that since Gutzwiller factors flatten the band, the resulting gap-to-bandwidth ratio is enhanced.

The ansatz \eqref{yrz_ansatz} can be related to the matrix:
\begin{align}
\hat{H}^{\text{YRZ}}_{\text{eff.}}&=
\begin{pmatrix}
\xi^{g}_{\vec k} & \Delta^{\text{PG}}_{\vec k}(p)\\
\Delta^{\text{PG}}_{\vec k}(p) &\xi^0_{\vec k}
\end{pmatrix},
\end{align}
using the following Green's function matrix:
\begin{align}
\hat{G}_{\vec k}(\omega)&=
[\omega - \hat{H}^{\text{YRZ}}_{\text{eff.}}]^{-1},
\label{green_def}
\end{align}
the first element is the electron Green's function:
\begin{align}
[\hat{G}_{\vec k}(\omega)]_{11} = 
\frac{1}{\omega-\xi_{\vec k}(p)-\displaystyle{\frac{|\Delta^{\text{PG}}_{\vec k}(p)|^2}{\omega-\xi^{0}_{\vec k}(p)}}}.
\label{G11}
\end{align}
Compared with \eqref{yrz_ansatz}, the only missing parts are the Gutzwiller renormalization factor $g_t(p)$ accounting for the loss of quasiparticle coherence, and the associated incoherent part: $G^{\text{YRZ}}_{\vec k}(\omega)\equiv g_t(p)[\hat{G}_{\vec k}(\omega)]_{11}+G_{\text{inc.}}$. In Ref.~\onlinecite{storey_hall_2016}, Storey showed that including this renormalization was detrimental to the fit with experimental Hall coefficients and thus left it out of the analysis. We do the same here. 

\section{Methods}
\label{section:methods}
\noindent
We use an effective 2x2 Hamiltonian formalism \cite{norman_modeling_2007,smith_thermal_2010,leblanc_signatures_2014} to unify the AF \cite{storey_hall_2016}, sAF \cite{eberlein_fermi_2016}, and YRZ \cite{storey_hall_2016} models, with all differences summarized in Table~\ref{table_models}. The Hamiltonian is:
\begin{align}
H=\sum_{\vec k} \Psi^{\dagger}_{\vec k}\hat{H}_{\vec k}\Psi^{\phantom{\dagger}}_{\vec k}.
\end{align}
with the spinor \mbox{$\Psi^\dagger_{\vec k} = \begin{pmatrix}c^\dagger_{\vec k\uparrow}&d^\dagger_{\vec k}\end{pmatrix}$} and the matrix:
\begin{align}
\hat{H}_{\vec k}=
\begin{pmatrix}
\xi_{\vec k}(p)
& \Delta_{\vec k}(p)
\\
\Delta_{\vec k}(p)
&\xi^{d}_{\vec k}(p)
\end{pmatrix}.
\label{Hamiltonian}
\end{align}
In each model $c^\dagger_{\vec k\uparrow}$ creates a Bloch electron of momentum~$\hbar \vec k$ and spin up, and the sum over $\vec k$ spans the original Brillouin zone. However, all models have different operators~$d^\dagger_{\vec k}$ and different dispersions $\xi_{\vec k}(p)$ and $\xi^{d}_{\vec k}(p)$, as given in Table~\ref{table_models}.

\begin{figure}
\includegraphics{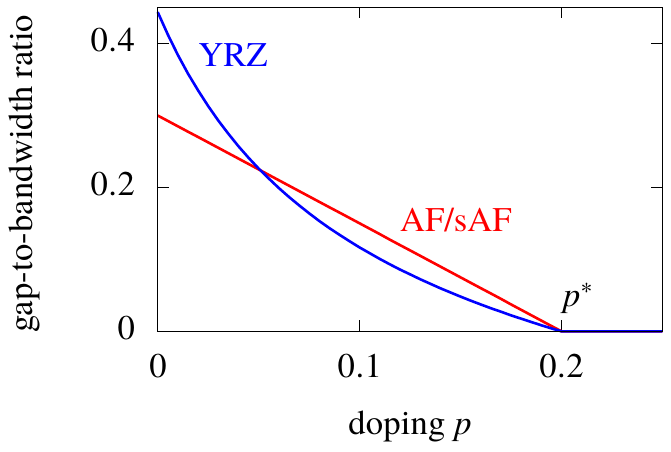}\hspace{10mm}
\caption{
Maximum gap-to-bandwidth ratio as a function of doping in the AF and sAF models (red) and in YRZ theory (blue). The ratio is computed as $\Delta_{(0,\pi)}(p)/(\xi_{(\pi,\pi)(p)}-\xi_{(0,0)}(p))$. The gap vanishes at $p^*=0.2$.
In the AF and sAF models, $\Delta_{\vec k}(p)=12t(p^*-p)$ and the bandwidth is $8t$, which yields a ratio of $0.3-1.5p$. 
In YRZ theory, the $d$-wave gap is maximum for $\vec k=(0,\pi)$ with a value of $\Delta_{(0,\pi)}(p)=3t(p^*-p)$ and the bandwidth changes with doping as $8t(\frac{2p}{1+p} + \frac{0.169}{(1+p)^2})$. The resulting gap-to-bandwidth ratio is thus $\frac{3}{8} (0.2-p) / (\frac{2p}{1+p}+\frac{0.169}{(1+p)^2})$.
}
\label{gap_ratios}
\end{figure}

Borrowing the idea from YRZ theory, each model's order parameter $\Delta_{\vec k}(p)$ vanishes at $p^*=0.2$: 
\begin{align}
\Delta_{\vec k}(p)=
\begin{cases}
\Delta_{\vec k}\cdot(p^*-p) &\text{for }0< p< p^*\\
0&\text{otherwise.}
\end{cases}
\label{gap}
\end{align}
All models have different $\Delta_{\vec k}$ given in Table~\ref{table_models}, in order to yield similar gap-to-bandwidth ratios as seen in Fig.~\ref{gap_ratios}. The $\vec k$-dependence of the YRZ gap have negligible effect on transport properties because it affects the low energy spectrum only around $(\pi,0)$ and~$(0,\pi)$.

\renewcommand{\arraystretch}{1.3}  
\begin{table}
\caption{Differences of the AF, sAF and YRZ models in the unified formalism.}
\label{table_models}
\begin{tabular}{| r | c  c  c |} 
\hline
& AF & sAF & YRZ
\\\hline
$\xi_{\vec k}(p)=$
& $\xi_{\vec k}=$ Eq. \eqref{band} & $\xi_{\vec k}=$ Eq. \eqref{band} & $\xi^{\text{g}}_{\vec k}(p)=$ Eq. \eqref{band_gutz}
\\
$\xi^{d}_{\vec k}(p)=$
& $\xi_{\vec k+\vec Q(p)}$ & $\xi_{\vec k+\vec Q(p)}$ & $\xi^{0}_{\vec k}(p)=$ Eq. \eqref{yrz_disp}
\\
$d^{\dagger}_{\vec k}=$
& $c^{\dagger}_{\vec k+\vec Q(p),\uparrow}$ & $c^{\dagger}_{\vec k+\vec Q(p),\downarrow}$ & ancillary
\\
$\vec Q(p)$=
& $(\pi,\pi)$ & $(\pi-2\pi p,\pi)$ & none
\\
$\Delta_{\vec k}$=
& $12t$ & $12t$ & $\frac{3}{2}t(\cos(k_x)-\cos(k_y))$
\\\hline
\end{tabular}
\end{table}

A $2\times2$ unitary transformation $\hat{U}_{\vec k}$ provides the eigenvalues $E_{n\vec k}=[ \hat{U}^{\dagger}_{\vec k} \hat{H}_{\vec k} \hat{U}^{\phantom{\dagger}}_{\vec k}]_{nn}$ for Hamiltonian~\eqref{Hamiltonian}. We let the doping dependence implicit from now on:
\begin{align}
E_{1(2),\vec k} &= \frac{\xi_{\vec k}+\xi^{d}_{\vec k}}{2}\mp\sqrt{\Big(\frac{\xi_{\vec k}-\xi^{d}_{\vec k}}{2}\Big)^2+\Delta_{\vec k}^2}.
\end{align}
The eigenstate quasiparticles are given by the operator $a^{\dagger}_{n\vec k}=[\hat{U}_{\vec k}^\dagger]_{n1} c^{\dagger}_{\vec k}+[\hat{U}_{\vec k}^\dagger]_{n2} d^{\dagger}_{\vec k}$, and the associated transformation matrix, analogous to a Bogoliubov transformation, can be written as:
\begin{align}
\hat{U}_{\vec k}&=
\begin{pmatrix}
\frac{\Delta_{\vec k}}{\sqrt{\Delta_{\vec k}^2 + (\xi_{\vec k} - E_{1\vec k})^2}}
&
\frac{\Delta_{\vec k}}{\sqrt{\Delta_{\vec k}^2 + (\xi_{\vec k} - E_{2\vec k})^2}}
\\
\frac{-\Delta_{\vec k}}{\sqrt{\Delta_{\vec k}^2 + (\xi_{\vec k} - E_{2\vec k})^2}}
&
\frac{\Delta_{\vec k}}{\sqrt{\Delta_{\vec k}^2 + (\xi_{\vec k} - E_{1\vec k})^2}}
\end{pmatrix}.
\end{align}

\subsection{Important quantities}
\label{section:relevantq}
\subsubsection{Velocity}
\noindent
As in Reference \onlinecite{storey_hall_2016,eberlein_fermi_2016}, velocities are chosen as:
\begin{align}
\vec v_{n\vec k}=\frac{1}{\hbar}\nabla_{\vec k}E_{n\vec k}
\label{velocity}
\end{align}
rather than $\frac{1}{\hbar}\vec \nabla_{\vec k}\xi_{\vec k}$. 
This choice for the velocity and its derivatives is subtle to justify rigorously~\cite{voruganti_conductivity_1992,paul_thermal_2003}\footnote{In particular, at $p^*$, the gap becomes arbitrarily small (because of Eq.~\ref{gap}) and electric breakdown should cause the failure of the semiclassical approximation \cite{ashcroft_solid_1976}.}. In Ref.~\cite{storey_hall_2016} it was shown that this choice is crucial to obtain agreement with experimental Hall coefficients~\cite{badoux_change_2016}.
In all models, the velocity of spin up electrons is $\vec v_{n\vec k,\uparrow}=\vec v_{n\vec k}$, but the one for spin down electrons depends on the model as given in Table~\ref{table_methods}.


\subsubsection{Spectral weight}
\label{section:spectral_weight}
\noindent
Since each eigenvalue has its velocity, our definitions for transport coefficients (section~\ref{subsec:transco}) require computing each eigenvalue's contribution to the spectral weight:
\begin{align}
[\hat{A}_{\vec k}(\omega)]_{ij}
&=
\sum_{n}
[\hat{U}_{\vec k}]_{in}[\hat{U}_{\vec k}^\dagger]_{nj}\frac{ \Gamma_{n\vec k} }{(\omega - E_{n\vec k})^2+\Gamma_{n\vec k}^{2}}
\label{spectralw2}
\\&\equiv 
\sum_{n}[\hat{A}_{n\vec k}(\omega)]_{ij}.
\label{bandspectralw}
\end{align}
Indices $i,j$ refer to the matrix element in the spinor basis, $n$ to the band index, and $\vec k$ covers the original Brillouin zone. In all models, the spectral weight for spin up electrons is given by $A_{n\vec k,\uparrow}=[\hat{A}_{n\vec k}(\omega)]_{11}$, but the one for spin down electrons depends on the model as given in Table~\ref{table_methods}.

\subsubsection{Scattering rates}
\label{section:rates}
\noindent
To stay in line with Refs.~\onlinecite{storey_hall_2016, eberlein_fermi_2016}, we consider the following two scattering rates:
\begin{align}
\text{constant-$\tau$ :}\qquad&
\Gamma_{n\vec k}=\frac{\hbar}{2\tau},
\\
\text{isotropic-$\ell$ :}\qquad&
\Gamma_{n\vec k}=\frac{at}{2\ell}\vert \vec v_{n\vec k}\vert+\zeta,
\label{scattering}
\end{align}
where $\zeta=10^{-5}$ prevents divergence of the spectral weight at saddle points of the dispersion. The same scattering rate is used for the two bands, \emph{i.e.} hole and electron pockets are always treated equivalently. The values used for $\tau$ and $\ell$, indicated in each figure's caption, were chosen as large as possible while ensuring successful numerical integration of transport coefficients.

\begin{figure}
\vspace{5mm}
\includegraphics{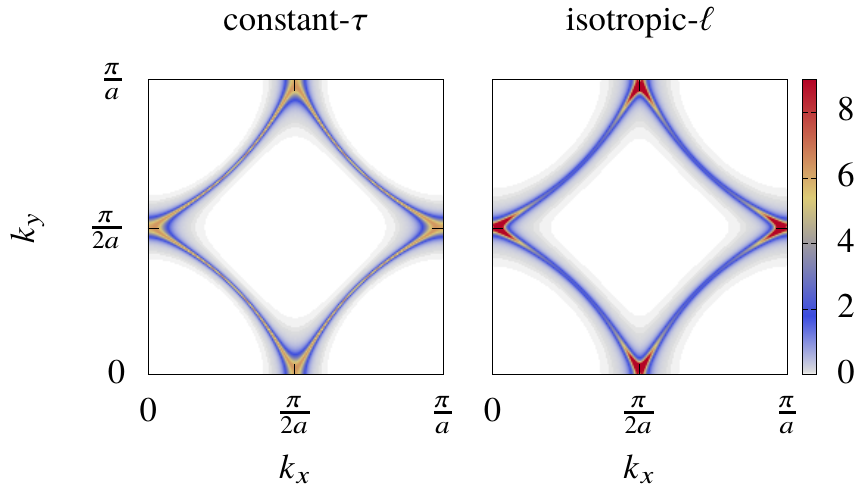}
\caption{Spectral weight at the Fermi level $A_{\vec k}(\epsilon=0)$ for band parameters $(t',t'')=(-0.17,0.05)t$, at doping $p=0.24$ (just passed the van Hove singularity, at $p_{\text{vHs}}=0.23$). The comparison of the constant lifetime $\tau$ approximation (left) and the isotropic mean-free path $\ell$ approximation (right) shows how isotropic-$\ell$ enhances the spectral weight near the antinode, compensating for the lower velocity of the states at the saddle points of the dispersion.}
\label{figure_normal_mdc}
\end{figure}

The main difference between the two approximations is that isotropic-$\ell$ enhances the weight of low velocity states as shown in Fig.~\ref{figure_normal_mdc}. Appendix~\ref{appendix:scattering} provides a complete comparison for the bare band case, also showing the effect of the van Hove singularity.

It was previously shown that experimental Seebeck coefficients~\cite{munakata_thermoelectric_1992, fujii_-plane_2002, kondo_contribution_2005} are more consistent with the isotropic-$\ell$ approximation~\cite{kondo_contribution_2005, storey_electron_2013} whereas experimental Hall coefficients~\cite{tsukada_negative_2006} are more consistent with a constant-$\tau$ approximation~\cite{narduzzo_violation_2008}. To remain general we compare both approximations throughout the rest of the paper. 

Although these two simple approximations have been widely used \cite{kondo_contribution_2005, storey_electron_2013, storey_hall_2016, abdel-jawad_anisotropic_2006, abdel-jawad_correlation_2007, kokalj_transport_2012}, experiments suggest alternative expressions for~$\Gamma_{n\vec k}$~\cite{hussey_phenomenology_2008}. For the Hall number in the AF model \cite{moutenet_phenomenological_nodate}, those alternative $\Gamma_{n\vec k}$ yield qualitative results equivalent to those of Ref.~\onlinecite{storey_hall_2016} and reproduced here. Clearly, more refined models of impurity scattering would be interesting~\cite{das_impurity_2016} in future studies.

\subsubsection{Doping}
\noindent
In the AF and sAF models, we find the chemical potential $\mu$ associated to a given doping $p$ with:
\begin{align}
p=1-\sum_{n}\int\frac{\text{d}^2 \vec k}{4\pi^2}f(E_{n\vec k}),
\label{nfermi}
\end{align}
where the eigenstates $E_{n\vec k}$ depend on $\mu$ implicitly.
At zero temperature, this is equivalent to Luttinger's rule
\begin{align}
p=1-\sum_{n}\int_{\text{Re}\{G_{n\vec k}(\omega=0))\}>0} \frac{\text{d}\vec k}{(2\pi)^2},
\label{luttinger_n}
\end{align}
with ${G}^{-1}_{n\vec k}(\omega)=\omega-E_{n\vec k}-i\Gamma_{n\vec k}$.

However, in YRZ theory, the quasiparticle $a_{n\vec k}^{\dag}$ associated with eigenvalue $E_{n\vec k}$ represents a mixture of an electron~$c_{\vec k}^{\dag}$ with the ancillary excitation represented by $d^\dagger_{\vec k}$ and thus Eq.~\eqref{nfermi} and~\eqref{luttinger_n} cannot be used. Instead it was prescribed{ \cite{yang_phenomenological_2006}} to count the electrons as follows:
\begin{align}
p^\text{YRZ}=1-2\int_{\text{Re}\{[\hat{G}_{\vec k}(\omega=0)]_{11})\}>0} \frac{\text{d}\vec k}{(2\pi)^2},
\label{luttinger}
\end{align}
where the Green's function Eq.~\eqref{G11} depends on $\mu$ implicitly. 
Strickly speaking, the doping computed for YRZ theory cannot be compared to that computed for the AF and sAF models; their natures are different: Eq.~\eqref{nfermi} and~\eqref{luttinger_n} count $1-\sum_{n\vec k}a_{n\vec k}^{\dag}a_{n\vec k}^{\phantom{\dag}}$ while Eq.~\eqref{luttinger} counts $1-2\sum_{\vec k}c_{\vec k\uparrow}^{\dag}c_{\vec k\uparrow}^{\phantom{\dag}}$, ignoring $\sum_{\vec k}d_{\vec k\uparrow}^{\dag}d_{\vec k\uparrow}^{\phantom{\dag}}$. It is nevertheless the accepted way to proceed \cite{yang_phenomenological_2006,storey_thermodynamic_2008, leblanc_specific_2009,storey_electron_2013,storey_hall_2016}.

\renewcommand{\arraystretch}{1.3}  
\begin{table}
\caption{Differences of the AF, sAF and YRZ models regarding the definitions of quantities relevant to transport. For each model $\vec v_{n\vec k\uparrow}=\vec v_{n\vec k}$ and $A_{n\vec k\uparrow}=[\hat{A}_{n\vec k}]_{11}$ from Eqs.~\eqref{velocity} and \eqref{bandspectralw}, respectively.}
\label{table_methods}
\begin{tabular}{| r | c  c  c |} 
\hline
& AF & sAF & YRZ
\\\hline
$\vec v_{n\vec k\downarrow}=$
& $\vec v_{n\vec k}$ & $\vec v_{n,\vec k-\vec Q(p)}$ & $\vec v_{n\vec k}$
\\
$A_{n\vec k\downarrow}=$
& $[\hat{A}_{n\vec k}]_{11}$ & $[\hat{A}_{n,\vec k-\vec Q(p)}]_{22}$ & $[\hat{A}_{n\vec k}]_{11}$
\\
$p=$
& Eq. \eqref{nfermi} & Eq. \eqref{nfermi} & Eq. \eqref{luttinger}
\\\hline
\end{tabular}
\end{table}

\subsection{Transport coefficients}
\label{subsec:transco}
\noindent
The AF, sAF and YRZ models we study are all formulated as $2\times2$ matrix models, so we define transport coefficients as sums on bands $n$ \cite{ashcroft_solid_1976}, each having respective  energy $E_{n\vec k}$, velocity $\vec v_{n\vec k}=-\frac{1}{\hbar}\nabla_{\vec k}E_{n\vec k}$ and scattering rate~$\Gamma_{n\vec k}$. 

\subsubsection{Hall effect}
\noindent With the electron charge $-e$ and the normalization volume $V$, the Hall number $n_H$ and resistivity $R_H$ are \cite{voruganti_conductivity_1992}:
\begin{align}
n_H=\frac{V}{eR_H},\quad R_H=\frac{\sigma_{xy}}{\sigma_{xx}\sigma_{yy}}.
\label{hall_def}
\end{align}
Given the Fermi-Dirac distribution, $f(\epsilon)=(\text{e}^{\beta\epsilon} +1)^{-1}$, conductivities $\sigma_{xx}$, $\sigma_{yy}$ and $\sigma_{xy}$ are expressed as:
\begin{align}
\sigma_{ab}&= \int\text{d}\epsilon\bigg(-\frac{\partial f(\epsilon)}{\partial \epsilon}\bigg) \sigma_{ab}(\epsilon),
\label{cond_integral}
\end{align}
with:
\begin{align} 
\sigma_{xx}(\epsilon)
&=
\frac{e^2\pi\hbar}{V}
\sum_{n\mathbf k\sigma}
v_{x,n\vec k\sigma}^2
A_{n\vec k\sigma}^2(\epsilon),
\label{longitudinal_integrand}
\\
\sigma_{xy}(\epsilon)&=-\frac{e^3(\pi\hbar)^2}{3V}\sum_{n\vec k\sigma} 
\bigg[
v_{x,n\vec k\sigma}^2
\frac{\partial v_{y,n\vec k\sigma}}{\partial k_y}+ 
v_{y,n\vec k\sigma}^2
\frac{\partial v_{x,n\vec k\sigma}}{\partial k_x}
\nonumber\\
&\phantom{=\frac{e^3}{3V}\sum_{n\vec k\sigma}}
-2 
v_{x,n\vec k\sigma}
v_{y,n\vec k\sigma}
\frac{\partial v_{x,n\vec k\sigma}}{\partial k_y}
\bigg]A_{n\vec k\sigma}^3(\epsilon),
\label{hall_integrand}
\end{align}
with a form equivalent to Eq.~\eqref{longitudinal_integrand} for $\sigma_{yy}$.
Those expressions are general enough to treat any scattering rate approximations through the spectral weight, along with the $x$-$y$ asymmetry of the sAF model. 
In particular, Eq.~\eqref{hall_integrand} is the anti-symmetrized version of Eq.~(1.25) of Ref.~\onlinecite{voruganti_conductivity_1992}, corresponding to $(\sigma_{H}^{xyz}-\sigma_{H}^{yxz})/2$ in their notation. It is necessary to use the antisymmetric form, like in experiments, because the combination of $x$-$y$ asymmetry and isotropic-$\ell$ approximation leads to slight quantitative differences for $\sigma_{H}^{xyz}$ and $-\sigma_{H}^{yxz}$.
Possible vertex corrections (scattering-in terms in the Boltzmann formalism) are  neglected here and for the Seebeck coefficient. 
The integral \eqref{cond_integral} can be evaluated exactly at zero temperature using $\lim_{T\rightarrow0}(-\frac{\partial f(\epsilon)}{\partial \epsilon}) = \delta(\epsilon)$.
Drude expressions~\mbox{$\sigma_{xx}=e^2\tau n/m^*$} and \mbox{$\sigma_{xy}=-e^3\tau^2n/m^*$} are only recovered in the parabolic band limit.

\subsubsection{Specific heat}
\noindent 
Each state of energy $\epsilon$ contributes an entropy \mbox{$S(\epsilon)=k_{B}(f(\epsilon)\ln f(\epsilon) + (1-f(\epsilon))\ln(1-f(\epsilon)))$} to the system. 
Taking the spectral weight into account is crucial~\cite{storey_thermodynamic_2008, leblanc_specific_2009}. This yields $C_{V}=T\frac{\partial S}{\partial T}$ as:
\begin{align}
C_{V}&=
\int\text{d}\epsilon \frac{\partial f(\epsilon)}{\partial T}\frac{\epsilon}{V}\sum_{n\vec k\sigma}A_{n\vec k\sigma}(\epsilon).
\end{align}
We are interested in $C_V/T$ at $T\rightarrow0$, which we compute with $\beta=500/t$, corresponding roughly to 6K (with $t=250$ meV). At such low temperature, the Sommerfeld expansion~\cite{ashcroft_solid_1976} shows that the expected result is proportional to the density of states at the Fermi level: $C_V/T=\pi^2k_B^2N(0)/3$, valid at least for non-singular integrand, \emph{i.e.} away from the doping $p_{\text{vHs}}$ where the van Hove singularity occurs.

\subsubsection{Seebeck coefficient}
\noindent For the Seebeck coefficient \cite{mcintosh_van_1996,kondo_contribution_2005,storey_electron_2013} we compute:
\begin{align}
S_{x}&=
\frac{1}{-eT}
\frac{\int\text{d}\epsilon\left(-\frac{\partial f(\epsilon)}{\partial \epsilon}\right)\epsilon\ \sigma_{xx}(\epsilon)
}{\int\text{d}\epsilon \left(-\frac{\partial f(\epsilon)}{\partial \epsilon}\right)\sigma_{xx}(\epsilon)}
,
\label{sbk}
\end{align}
with $\sigma_{xx}(\omega)$ given by~\eqref{longitudinal_integrand}, and with an analogous expression for $S_{y}$ to treat the $x$-$y$ asymmetry of the sAF model. Again, we are interested  in $S_x/T$ at $T\rightarrow0$, which we compute, in that case, with $\beta=100/t$, roughly equivalent to 30K. Higher temperatures must be used, compared to $C_V$, for the numerical integration to succeed, but we verified that the value of $S_x/T$ is stabilized at that temperature (except for the vicinity of the $p_{vHs}$). More details on this expression for the Seebeck coefficient can be found in Refs.~\onlinecite{mcintosh_van_1996,kondo_contribution_2005}.

\section{Results}\label{section:results}

\begin{figure}
\includegraphics{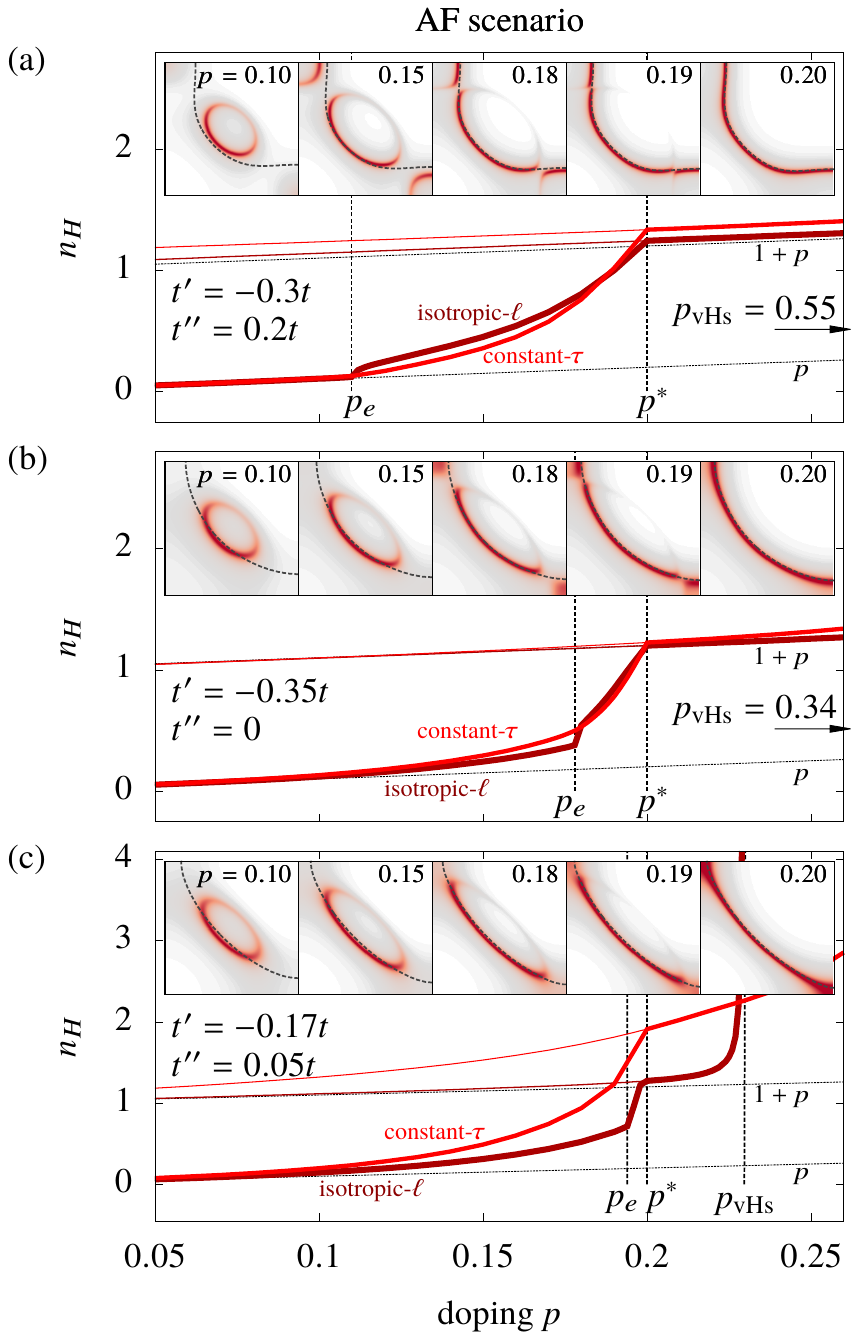}
\caption{Hall number ($n_H$) and Fermi surfaces ($\sum_{\sigma}A_{n\vec k\sigma}(\omega=0)$) in the antiferromagnetic model. The rise in Hall number starts at $p_e$, where the electron pockets at $(\pi,0)$ and $(0,\pi)$ appear at the Fermi surface, and ends at $p^*=0.2$, where the gap vanishes. Three sets of band parameters (a) $(t',t'')=(-0.3,0.2)t$, similar to YBCO, (b) $(t',t'')=(-0.35,0)t$, similar to BSCCO, and (c) $(t',t'')=(-0.17,0.05)t$ similar to LSCO, yield different values of~$p_e$. The thick lines are the AF results and the thin lines are the bare band results; they merge together above $p^*=0.2$. Constant lifetime $\tau=5\hbar/t$ and isotropic mean-free path $\ell=10a$ approximations are identified. The dotted lines are guides to the eye following $p$ and $1+p$, while dashed lines identify $p_e$, $p^*$, and the doping $p_{\text{vHs}}$ where the van Hove singularity occurs. The latter is in the plot range only in case (c).}
\label{figure_hall}
\end{figure}

\noindent
Results are very similar for all three models.
In all of them, at low doping, small hole pockets around $(\pm\pi/2,\pm\pi/2)$ are the only contributors to transport. At a doping $p_{e}$, electron pockets appear around $(\pm\pi,0)$ and $(0,\pm\pi)$, and when the gap closes at $p^*$, they reconnect with the $(\pm\pi/2,\pm\pi/2)$ hole pockets to recover the single large Fermi surface of the bare band~$\xi_{\vec k}$.

Changing band parameters has equivalent consequences in every model studied; it changes the position of $p_e$. Thus, to lighten this section, we take the AF results as a reference to compare the sAF and YRZ results. Section \ref{section:af} presents the AF results for various band structures and highlights general observations that are representative of all three models. The sAF results and the YRZ results follow in section \ref{section:saf} and \ref{section:yrz} respectively, using only one set of band parameters for each to highlight the differences with the AF results.

\subsection{Antiferromagnetism}
\label{section:af}

\noindent
Fig.~\ref{figure_hall} shows the Fermi surfaces across the $p^*$ transition in the AF model, and the rises in Hall number from $p$ to $1+p$ for three different sets of band parameters and the two scattering rate approximations.  

For every set of band parameters, the appearance of the electron pockets at $p_{e}$ marks the beginning of a \emph{progressive} rise of the Hall number ending at~$p^*$, as was studied in detail in Ref.~\onlinecite{storey_hall_2016}.

For the same gap amplitude $\Delta_{\vec k}(p)=12t(0.2-p)$, different band parameters yield different values of $p_e$.  
For example, in Fig.~\ref{figure_hall}(a), the electron pockets appear at $p=0.11$, while for the band parameters of Fig.~\ref{figure_hall}(c), they appear at $p_e=0.195$. 
As one can see, the closer $p_{\text{vHs}}$ is above~$p^*$ the closer~$p_e$ is below~$p^*$.
Indeed, since the electron pocket forms from the bare band near $(\pi,0)$, the closer the bare band is to $(\pi,0)$ the smaller the electron pocket is and the quicker it vanishes with the gap.

While the Hall numbers of Fig.~\ref{figure_hall} all follow closely $n_H=p$ at low doping, for high dopings only the isotropic-$\ell$ approximation yields values that follow closely $n_H=1+p$ beyond $p^*$. The constant-$\tau$ approximation usually yields values exceeding $1+p$, due to the ellipticity of the electron pockets \cite{chatterjee_fractionalized_2016} and the $\vec k$-dependence of the velocity~\cite{zou_theory_2017}. In the isotropic-$\ell$ approximation, the spectral weight compensates those effects to give precisely $1+p$ (see appendix~\ref{appendix:scattering}). Note that experimental values for the Hall resistivity exceed $n_H=1+p$ \cite{tsukada_negative_2006}, agreeing better with the constant-$\tau$ approximation.

\begin{figure}
\includegraphics{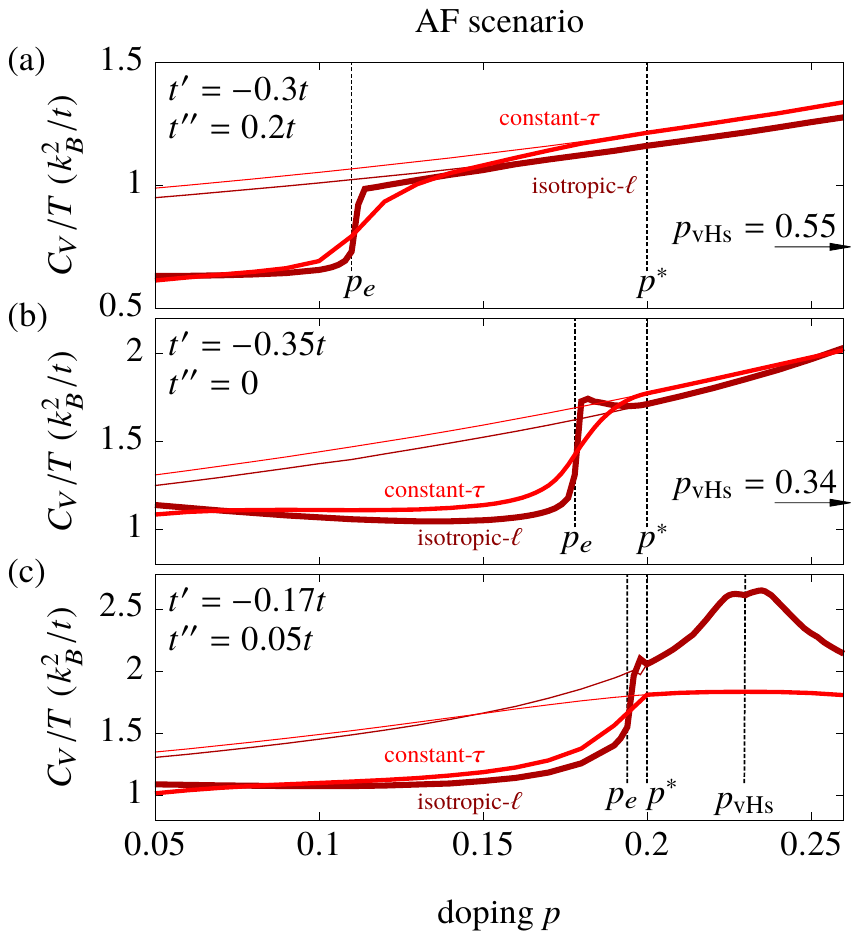}
\caption{Specific heat ($C_V/T$) in the antiferromagnetic model, at temperature $\beta=500/t$, equivalent to $T\approx6$K, shown as a function of doping for band parameters (a) $(t',t'')=(-0.3,0.2)t$, (b) $(t',t'')=(-0.35,0)t$, and (c) $(t',t'')=(-0.17,0.05)t$. The corresponding Fermi-surfaces were shown in Fig.~\ref{figure_hall}. Labels identify constant lifetime $\tau=5\hbar/t$ and isotropic mean-free path $\ell=10a$ approximations, along with the doping $p_e$ where electron pockets appear at the Fermi surface, the doping $p^*$ where the gap vanishes, and the doping $p_{\text{vHs}}$ where the van Hove singularity occurs. The thick lines are the AF results and the thin lines are the bare band results; they merge together above $p^*=0.2$.}
\label{figure_specific_heat}
\end{figure}

Fig.~\ref{figure_specific_heat} shows the specific heat across the $p^*$ transition in the AF model, for the same three sets of band parameters and the two scattering rate approximations. The doping
$p_{e}$ is marked by a rapid rise in $C_V/T$, corresponding to the gain in density of states associated with the electron pockets appearing at the Fermi surface.
As a consequence, the position of the rise is locked to $p_e$, and completely independent of $p^*$, strongly contrasting with the rise in Hall number.

\begin{figure}
\includegraphics{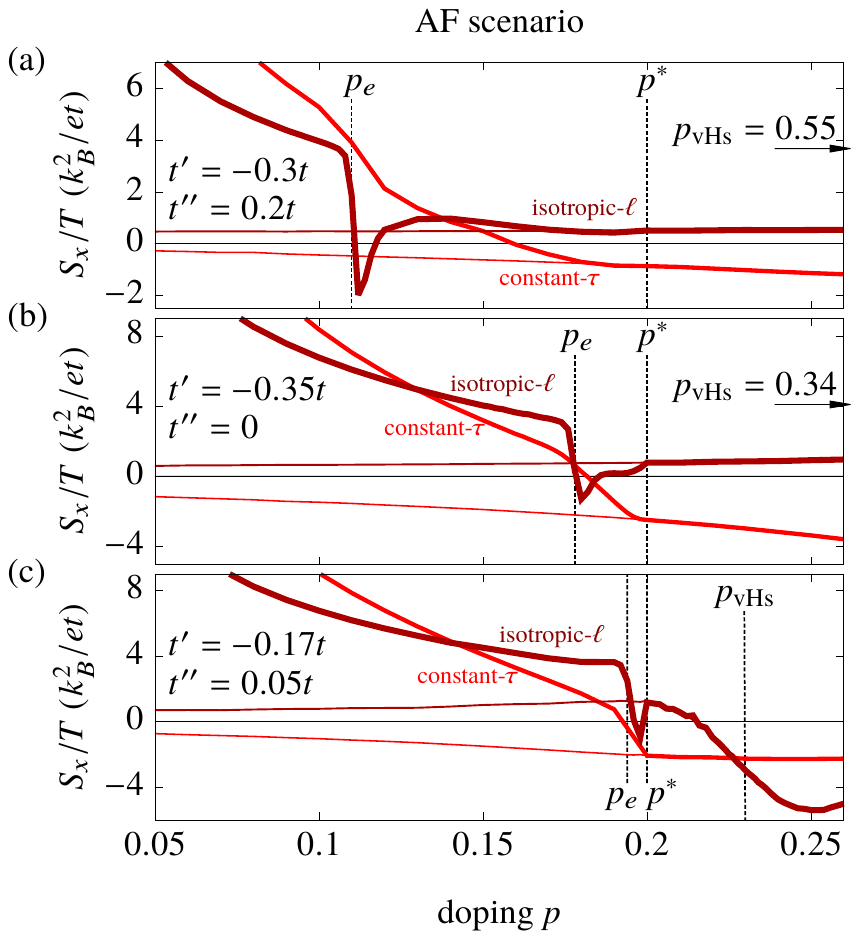}
\caption{
Seebeck coefficient ($S_x/T$) in the antiferromagnetic model, at temperature $\beta=100/t$, equivalent to $T\approx30$K, shown as a function of doping for band parameters (a) $(t',t'')=(-0.3,0.2)t$, (b) $(t',t'')=(-0.35,0)t$, and (c) $(t',t'')=(-0.17,0.05)t$. The corresponding Fermi-surfaces were shown in Fig.~\ref{figure_hall}. Labels identify constant lifetime $\tau=5\hbar/t$ and isotropic mean-free path $\ell=10a$ approximations, along with the doping $p_e$ where electron pockets appear at the Fermi surface, the doping $p^*$ where the gap vanishes, and the doping $p_{\text{vHs}}$ where the van Hove singularity occurs. The thick lines are the AF results and the thin lines are the bare band results; they merge together above $p^*=0.2$.}
\label{figure_seebeck}
\end{figure}

The rises of Fig.~\ref{figure_specific_heat} are sharper in the isotropic-$\ell$ approximation than in the constant-$\tau$ approximation. This is because the electron pockets correspond to the minima of band $E_{2,\vec k}$ and therefore their velocity vanishes, $\frac{1}{\hbar}\vec \nabla_{\vec k}E_{2,\vec k}=0$, at the doping where they first appear. Since the broadening is proportional to the velocity in the isotropic-$\ell$ approximation, this vanishing causes a sharp spectral weight at the Fermi level and consequently a very sharp jump in the density of states when electron pockets appear at the Fermi surface, resulting in the rise of $C_V/T$.

Fig.~\ref{figure_seebeck} shows the Seebeck coefficient across the $p^*$ transition in the AF model, again for the three sets of band parameters and the two scattering rate approximations. Contrary to the Hall number and the specific heat, the transition is accompanied not by a rise, but by a drop in $S_x/T$; the results for $p<p_e$ are a lot higher than those for $p>p^*$.
Furthermore, the progression from Fig~\ref{figure_seebeck}(a) to Fig~\ref{figure_seebeck}(c) indicates that this drop is locked to $p_e$ rather than to $p^*$.


In the isotropic-$\ell$ approximation, the Seebeck coefficients fall sharply to negative values at $p_e$ before returning to the bare band positive values. This is because the Seebeck coefficient is sensitive to particle-hole asymmetry, hence electron pockets introduce a negative contribution to~$S_x/T$.
In the constant-$\tau$ approximation, the low velocity of electron pockets reduces this negative contribution, so the drops of Fig.~\ref{figure_seebeck} are not sharp but progressive, continuing down to the bare band negative values. As already mentioned, the experimental Seebeck coefficients~\cite{munakata_thermoelectric_1992, fujii_-plane_2002, kondo_contribution_2005} typically agree better with the positive values of the isotropic-$\ell$ approximation~\cite{kondo_contribution_2005, storey_electron_2013}, contrary to the Hall number.

Lastly, as one can see in Fig.~\ref{figure_seebeck}(c) and subsequent Seebeck results, the aforementioned numerical difficulties associated with Eq.~\eqref{sbk} cause some noise in the results.

\subsection{Spiral antiferromagnetism}
\label{section:saf}
\begin{figure}
\includegraphics{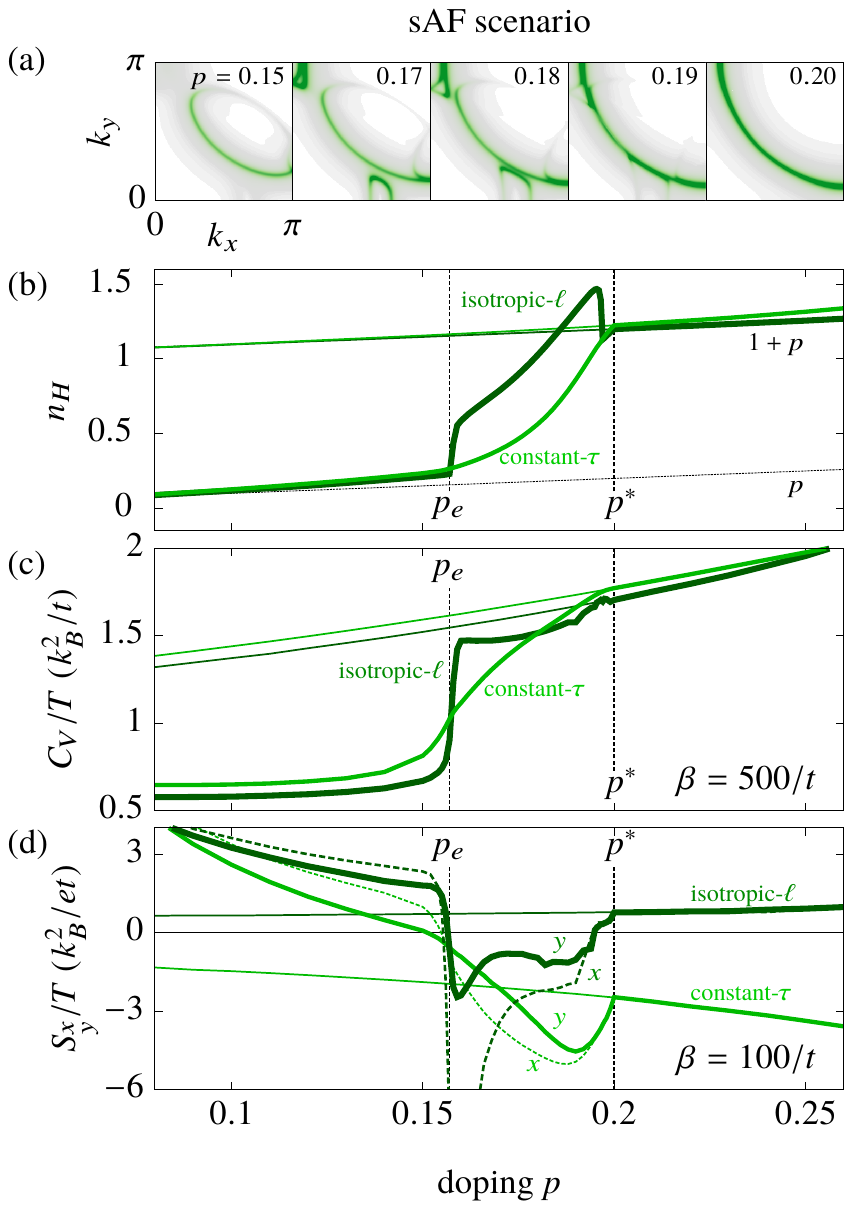}
\caption{Hall number ($n_H$), specific heat ($C_V/T$), and Seebeck coefficient ($S_x/T$ and $S_y/T$) in the incommensurate spiral antiferromagnetic model. Band parameters are $(t',t'')=(-0.35,0)t$. (a) The Fermi surfaces, (b) the Hall number $n_H$ at $T\rightarrow0$, (c) the specific heat $C_V/T$ at $\beta=500/t$ ($T\approx6$K) and (d) the Seebeck coefficient $S_x/T$ at $\beta=100/t$ ($T\approx30$K) are all shown as a function of doping $p$.
Labels identify constant lifetime $\tau=5\hbar/t$ and isotropic mean-free path $\ell=10a$ approximations, along with the doping $p_e$ where electron pockets appear at the Fermi surface, and the doping $p^*$ where the gap vanishes. The thick lines are the sAF results and the thin lines are the bare band results; they merge together above $p^*=0.2$. Labels $x$ and $y$ identify which lines are the $S_x/T$ and $S_y/T$ for the Seebeck coefficients.}
\label{figure_comp_sAF}
\end{figure}

\noindent
Fig.~\ref{figure_comp_sAF} shows the signatures of the $p^*$ transition for the sAF model for only one set of band parameters, $(t',t'')=(-0.35,0)t$. Varying the band parameters leads to the same observations as in the previous section for the AF model, so in what follows we only highlight the main differences between the sAF and the AF results already shown.

Most differences come from the Fermi surfaces, shown respectively in Fig.~\ref{figure_comp_sAF}(a) and Fig.~\ref{figure_hall}(b). Because of the wave vector $(\pi-2\pi p,\pi)$ of the sAF model, all pockets are slightly displaced along $x$ and the pocket around $(\pi,0)$ has a different shape from that at $(0,\pi)$. Ref.~\onlinecite{eberlein_fermi_2016} gives a complete view of these Fermi surfaces.
Moreover, the velocity of the hole pocket is higher, relative to that of the electron pocket, in the sAF model than in the AF model (not shown). These differences in Fermi surfaces and in velocities are the main factor causing differences in transport coefficients.

In Fig.~\ref{figure_comp_sAF}(b), the Hall number of the isotropic-$\ell$ approximation deviates substantially from the monotonic rise of the constant-$\tau$ case studied in Ref.~\onlinecite{eberlein_fermi_2016}. This deviation comes from a reduction of $\sigma_{xy}$ due to the aforementioned lower velocity of the electron pockets. Since the isotropic-$\ell$ approximation enhances low velocity states, this detrimental contribution appears more clearly than in the constant-$\tau$ approximation. In other words, the less the electron pockets contribute, as in the constant-$\tau$ approximation, the better the agreement with experiments~\cite{badoux_change_2016}. 



Lastly, the Seebeck coefficients $S_x/T$ and $S_y/T$ in Fig.~\ref{figure_comp_sAF}(d) undergo strong variations in the presence of the sAF electron pockets, stronger for $S_x/T$ than for $S_y/T$. Even in the constant-$\tau$ approximation, with a diminished electron-pocket contribution, $S_x/T$ and $S_y/T$ have minima between $p_e$ and $p^*$, contrasting the monotonic decrease of the corresponding AF results in Fig.~\ref{figure_seebeck}(c). Again, we remind the reader that the experimental Seebeck coefficient~\cite{munakata_thermoelectric_1992, fujii_-plane_2002, kondo_contribution_2005} agrees better with the isotropic-$\ell$ approximation \cite{kondo_contribution_2005, storey_electron_2013}. In the latter case, the AF model and the sAF model display very similar dips, except that they are much deeper in the sAF case, and display enhanced substructures, with a larger range of negative values.

\begin{figure}
\includegraphics{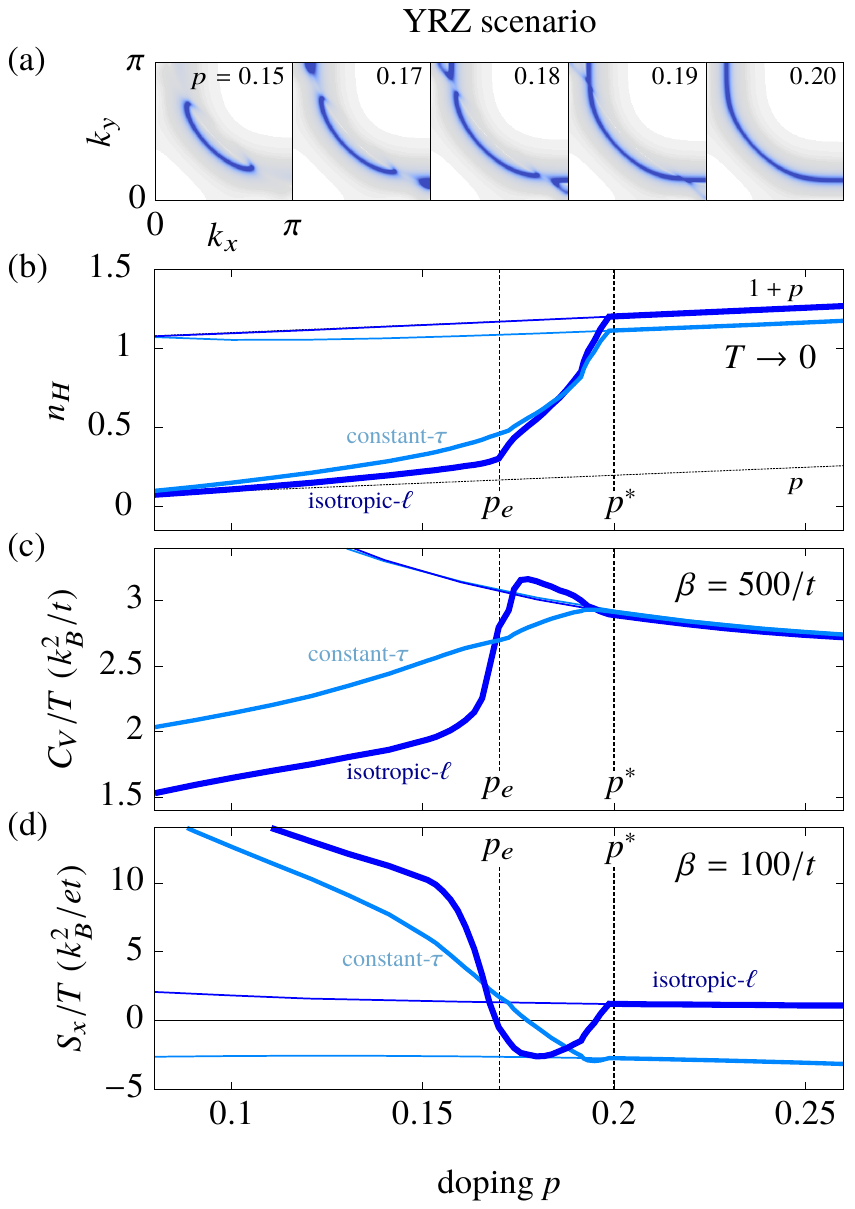}
\caption{
Hall number ($n_H$), specific heat ($C_V/T$), and Seebeck coefficient ($S_x/T$) in the Yang-Rice-Zhang theory. 
The usual YRZ band parameters $(t',t'')=(-0.3,0.2)t$ are renormalized by Gutzwiller factors, so even bare band results (thin lines) are different from the ones obtained in the AF and sAF models. 
(a) The Fermi surfaces, (b) the Hall number $n_H$ at $T\rightarrow0$, (c) the specific heat $C_V/T$ at $\beta=500/t$ ($T\approx6$K) and (d) the Seebeck coefficient $S_x/T$ at $\beta=100/t$ ($T\approx30$K) are all shown as a function of doping $p$.
Labels identify constant lifetime $\tau=5\hbar/t$ and isotropic mean-free path $\ell=10a$ approximations, along with the doping $p_e$ where electron pockets appear at the Fermi surface, the doping $p^*$ where the gap vanishes. The thick lines are the sAF results and the thin lines are the bare band results; they merge together above $p^*=0.2$.}
\label{figure_comp_YRZ}
\end{figure}

\subsection{Yang-Rice-Zhang theory}
\label{section:yrz}

Fig. \ref{figure_comp_YRZ} shows the signatures obtained across the $p^*$ transition in YRZ theory. The usual band parameters for YRZ theory, $(t',t'')=(-0.3,0.2)t$, are strongly renormalized by Gutzwiller factors, which makes them more comparable to non-renormalized $(t',t'')=(-0.35,0)t$ than to non-renormalized $(t',t'')=(-0.3,0.2)t$. Nevertheless, renormalized bands are very different; even the bare band results of this section differ from those of the two previous sections. Nevertheless, the qualitative observations highlighted in our analysis of the AF model holds in YRZ theory; this section focuses on the differences between both models.

Differences with the AF results are not all explained by the Fermi surfaces of Fig.~\ref{figure_comp_YRZ}(a). Actually, those are so similar to that of the AF model that most differences are caused by the Gutzwiller factors in the dispersion. These factors cause a chain of effects, of which the most relevant are: (i) a reduction of the bandwidth, which comes with (ii) a decreased velocity, (iii) an increased density of states at given energy or doping, and (iv) a relative broadening of band edges as a function of energy or doping.

Fig.~\ref{figure_comp_YRZ}(b) shows that the Hall number is almost indistinguishable from the one obtained in the AF model in Fig.~\ref{figure_hall}(b), as studied in Ref.~\onlinecite{storey_hall_2016}. This holds in both scattering approximations, and is consistent with the very similar Fermi surfaces of the two models.


Lastly, in Fig.~\ref{figure_comp_YRZ}(c), the rise in $C_V/T$ is broader than in the AF case; in fact, for the constant-$\tau$ approximation, it is more a change of slope than a sharp rise. Accordingly, in Fig.~\ref{figure_comp_YRZ}(d), the drop in $S_x/T$ is broader than in the AF and the sAF models, such that it becomes negative in the whole interval $p_e$ to $p^*$ in the isotropic-$\ell$ approximation.
This broadening is consistent with the flattening of the band due to Gutzwiller factors; the upper band edge, \emph{i.e.} the bottom of the electron pocket, represents a larger fraction of the overall bandwidth. Moreover, in the constant-$\tau$ approximation, the reduced velocity of electron pockets is further decreased by Gutzwiller factors, making their contributions almost invisible in YRZ transport results.


\section{Conclusion}\label{section:conclusion}
\noindent
We studied the transport signatures of the low temperature transition at $p^*$ in three models relevant to hole-doped cuprates: the AF model, sAF model, and YRZ theory. We found that, together with the rise of the Hall number~$n_H$ studied previously \cite{storey_hall_2016, eberlein_fermi_2016}, all studied models predict a rise in specific heat $C_V/T$ and a drop in the Seebeck coefficient $S_x/T$ as doping increases.

Our results are consistent with the known trends of experiments~\cite{loram_condensation_2000,loram_specific_1998,momono_low-temperature_1994,mcintosh_van_1996,yoshida_low-energy_2007,kondo_contribution_2005, munakata_thermoelectric_1992, fujii_-plane_2002, daou_thermopower_2009}. Finite temperature specific heat measurements indicate that the low temperature density of states increases significantly with doping \cite{loram_condensation_2000,loram_specific_1998,momono_low-temperature_1994}, 
and finite temperature Seebeck coefficients indicate that the low-temperature $S_x/T$ decreases significantly with doping \cite{mcintosh_van_1996,kondo_contribution_2005, munakata_thermoelectric_1992, fujii_-plane_2002, daou_thermopower_2009}. However, the normal state doping dependence of those probes is not well documented. Low-temperature measurements with superconductivity suppressed and a well resolved $p^*$ remain to be published. The comparison of such experiments with our prediction of a rapid rise in specific heat $C_V/T$ and a drop in the Seebeck coefficient $S_x/T$ will be a stringent test for the assumptions underlying the AF, sAF and YRZ models studied here, all of which fully explain the rapid rise in Hall number \cite{badoux_change_2016,storey_hall_2016,eberlein_fermi_2016}.

To a great extent, the positions in doping of the $C_V/T$ rise, and $S_x/T$ drop are controlled by $p_e$, the doping at which electron pockets appear at the Fermi surface before $p^*$ in all considered models. With the AF model, we showed that this doping $p_e$ depends on band structure. From our results, we can infer that the closer $p_{\text{vHs}}$ is above $p^*$, the closer~$p_e$ is below~$p^*$.
This indicates that not all compounds may provide equivalent evidence of this separation of $p_e$ and $p^*$. For example, LSCO, known to have $p^*$ very close to $p_{\text{vHs}}$ might not be a good candidate, single layer Hg or Tl compounds being preferable options. However, this remark holds given the same gap amplitude $\Delta_{\vec k}(p)$ for all band structures. Recent experiments on LSCO presented a rise of the Hall number spanning a finite range of doping~\cite{laliberte_origin_2016}, consistent with an appreciable separation of $p_e$ and $p^*$. A smaller gap amplitude in LSCO, consistent with its lower $T^*$, may be the explanation; it would let the electron pockets survive on a larger doping range~\cite{storey_hall_2016,eberlein_fermi_2016}.

The electron pockets at $(\pi,0)$ and $(0,\pi)$ in the transition regime $p_e<p<p^*$ provide model-specific predictions.
In that range of doping, we studied carefully the distinctive signatures of each model both in the isotropic mean-free path $\ell$ and the constant lifetime $\tau$ approximations, assuming that the electron pockets have the same mean-free path or constant lifetime as the hole pocket. Looking at the results, the only systematic discriminating signature comes from the Seebeck coefficient in the isotropic-$\ell$ approximation. The prediction consists of a dip of $S_x/T$ to negative values between $p_e$ and $p^*$, and with a characteristic shape in each model. In the AF model, the dip is negative only within a narrow doping range; in the sAF model, the dip is negative for a larger range of doping and displays characteristic substructures; and in YRZ theory, the dip is round and broad.
Those signatures may change at lower temperatures; the finite temperature $\beta=100/t$ used in our computations of $S_x$, equivalent roughly to $T=30K$, gives a good indication of the $T=0$ limit, except near the van Hove singularity. The analysis of the temperature dependence is beyond the scope of this work and can be found for YRZ theory in Ref.~\onlinecite{leblanc_specific_2009,storey_electron_2013,storey_hall_2016}.

In the end, what this work really shows is how the AF, sAF and YRZ models all predict two separate dopings marking the $p^*$ transition in transport properties. The rise in specific heat and drop in Seebeck coefficient should be found, not at $p^*$ like for the rise in the Hall number, but at a separate doping $p_e$. This separation is caused by the characteristic electron pockets of these models.
To this day, no such electron pocket has been reported in photoemission or quantum oscillations experiments near $p^*$. Electron pockets of a different nature are present~\cite{Harrison_e-pockets_2011,Doiron:2007} in the charge-density wave regime~\cite{chang_direct_2012}, which is not considered here. The separation predicted here for $p_e$ and $p^*$ in transport properties offers a new way to address this question experimentally.
The absence of such a separation would raise very serious doubts on any theory relying on $(\pi,0)$ and $(0,\pi)$ electron pockets to close the pseudogap at $p^*$.
\footnote{
FL$^*$ theory provide some interesting examples.
The U(1) algebraic charge liquid of Ref.~\onlinecite{qi_effective_2010} leads to an effective Hamiltonian very similar to the AF and YRZ models treated here and so we believe conclusions presented here for the AF model should hold in this FL$^*$ theory. However, the $\mathds Z_2$ FL$^*$ theory of Ref.~\onlinecite{chatterjee_fractionalized_2016} leads to an effective spectrum which does not rely on $(\pi,0)$ and $(0,\pi)$ electron pockets, but rather on large spinon pockets, so the results may differ from ours.  }

There exist theories of the pseudogap without the electron pockets studied here. The SU(2) phenomenological theory~\cite{montiel_effective_2017,morice_evolution_2017} was recently shown to agree with the rise in Hall number, but its pseudogap is particle-hole symmetric and without electron pockets. Alternatively, strongly correlated-electron methods for hole-doped cuprates can obtain particle-hole asymmetric pseudogaps~\cite{SordiChargeTransfer:2016,MacridinPseudogap:2006} without electron pockets. For example, very clear Fermi arcs without broken symmetry \cite{Senechal:2004,Civelli:2005,kyung_pseudogap_2006,stanescu_fermi_2006,Haule:2007,Kancharla:2008,sakai_evolution_2009,Ferrero:2009,FerreroEPL:2009,GullFerrero:2010} are obtained. In this case, the spectral weight gapped at $(0,\pi)$ and $(\pi,0)$ is strongly incoherent and does not form a Fermi surface. In these methods, vertex corrections and the effect of elastic scattering off impurities will need to be included to make reliable predictions in the regime of interest here.

\section{Acknowledgments}
We acknowledge S.~Badoux, S.~Chatterjee, N.~Doiron-Leyraud, C.-D.~H\'ebert, F.~Lalibert\'e, B.~Michon, J.~Leblanc, R.~Nourafkan, C.~Proust, G.~Sordi, J.~Storey, and L.~Taillefer for fruitful discussions. We also acknowledge A.~Moutenet and A.~Georges for sharing results and ideas with us. This work was partially supported by the Natural Sciences and Engineering Research Council (Canada) under grant RGPIN-2014-04584, the Fonds Nature et Technologie (Qu\'ebec) and the Research Chair on the Theory of Quantum Materials (A.-M.S.T.).

\appendix
\section{Isotropic-$\ell$ vs constant-$\tau$}
\label{appendix:scattering}

\begin{figure}
\includegraphics{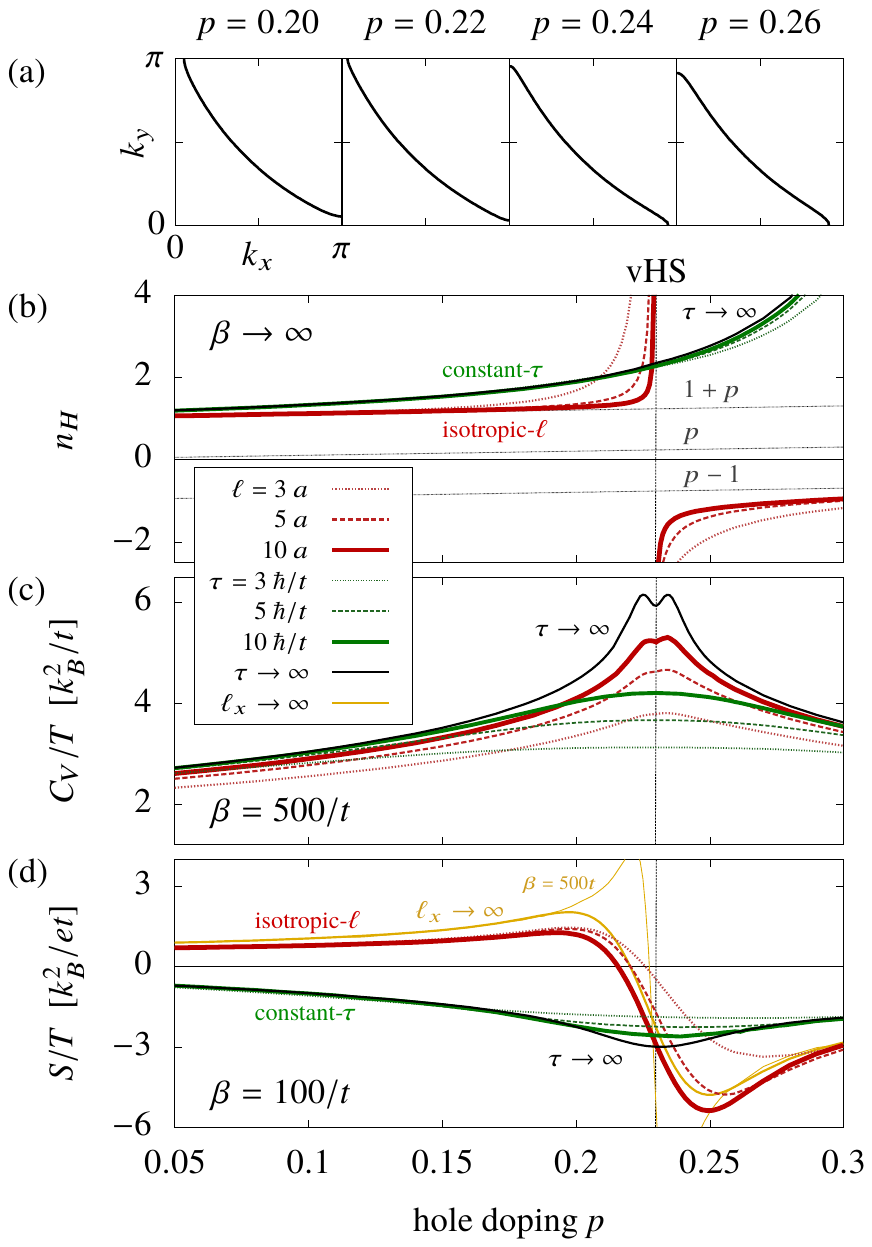}
\caption{Results for the bare band with band parameters $(t',t'')=(-0.17,0.05)t$. (a) Fermi surfaces ($\xi_{\vec k}=0$) as a function of doping, two on each side of the van Hove singularity (vHs) at $p=0.23$. (b-d) Comparison of (b) Hall numbers $n_H=1/eR_H$ in the limit of zero temperature, (c) specific heats $C_V/T$ at $\beta=500/t$, and (d) Seebeck coefficients $S(T)/T$ at $\beta=100/t$, as a function of doping. The isotropic mean-free path $\ell$ approximation is in red, while the constant lifetime $\tau$ approximation is in green and the legend identify the values of $\tau$ and $\ell$ used. $\tau\rightarrow\infty$ denotes the standard Boltzmann transport theory results, and $\ell_x\rightarrow\infty$ the constant mean-free path approximation used in References \onlinecite{kondo_contribution_2005, storey_electron_2013}, for comparison. In the latter case, it was possible to compute $S/T$ at $\beta=500/t$, revealing the divergence of the Seebeck coefficient at the van~Hove singularity in the constant-$\ell$ approximation for lower temperatures.}
\label{figure_normal_transport}
\end{figure}

\noindent
Using the constant-$\tau$ approximation of section~\ref{section:rates} is not the same as using usual Boltzmann transport theory.
The two are only equivalent for~$\tau\rightarrow\infty$.
In the usual Boltzmann theory \cite{ashcroft_solid_1976}, $\tau$
cancels out of ratios like the Hall resistivity and the Seebeck coefficient, so its actual value has no importance.
In the spectral weight formulation of section~\ref{subsec:transco}, a finite value of $\tau$ appears as a Lorentzian broadening of the spectral weight, and it cannot cancel out of those ratios.

To make a similar point, the isotropic-$\ell$ approximation of section~\ref{section:rates} can be compared to the constant mean-free-path approximation used in Ref.~\onlinecite{kondo_contribution_2005, storey_electron_2013}. In these works, it is $v_{x,n\vec{k}}\tau$ which is assumed constant and cancels out of the Seebeck ratio. We would therefore identify this approximation as $\ell_x\rightarrow\infty$. However, this approximation is incompatible with the Hall coefficient because both $v_{x}$ and $v_{y}$ enter the expression of $\sigma_{xy}$.

To illustrate the differences between all the approximations discussed, Fig.~\ref{figure_normal_transport} shows the bare band results, \emph{i.e.} $\Delta_{\vec k}(p)=0$, for various values of isotropic-$\ell$ and constant-$\tau$, along with the conventional Boltzmann theory result, denoted by $\tau\rightarrow\infty$ and the approximation of Refs.~\onlinecite{kondo_contribution_2005, storey_electron_2013}, denoted by $\ell_x\rightarrow\infty$.
In the isotropic-$\ell$ approximation, the Hall number and Seebeck coefficient both change sign precisely at the van Hove singularity at $p_{\text{vHs}}=0.23$. On the other hand, for the constant-$\tau$ approximation, the Hall coefficient changes sign much farther, beyond $p=0.30$, and the Seebeck coefficient changes sign a lot before, below $p=0.05$. Changing the value of $\tau$ and $\ell$ only causes broadening, the clearest case being the specific heat of Fig.~\ref{figure_normal_transport}(c) close to the van Hove singularity. 
All differences between the constant-$\tau$ and isotropic-$\ell$ approximations come from differences in the corresponding spectral weights of Fig.~\ref{figure_normal_mdc} which shows how the isotropic-$\ell$ approximation enhances the weight of low velocity of states near $(\pi,0)$ and $(0,\pi)$.


%

\end{document}